# Deciphering alloy composition in superconducting single-layer FeSe$_{1-x}$S$_x$ on SrTiO$_3$(001) substrates by machine learning of STM/S data

Qiang Zou[†], Basu Dev Oli[†], Huimin Zhang[†], Joseph Benigno[†], Xin Li[‡], and Lian Li[†,*]

[†]Department of Physics and Astronomy, West Virginia University, WV 26506, USA

[‡]Lane Department of Computer Science and Electrical Engineering, West Virginia University, WV 26506, USA

*To whom correspondence may be addressed. Email: lian.li@mail.wvu.edu

## Abstract

Scanning tunneling microscopy (STM) is a powerful technique for imaging atomic structure and inferring information on local elemental composition, chemical bonding, and electronic excitations. However, traditional methods of visual inspection can be challenging for such determination in multi-component alloys, particularly beyond the dilute limit due to chemical disorder and electronic inhomogeneity. One viable solution is to use machine learning to analyze STM data and identify patterns and correlations that may not be immediately apparent through visual inspection alone. Here, we apply this approach to determine the Se/S concentration in superconducting single-layer FeSe$_{1-x}$S$_x$ alloy epitaxially grown on SrTiO$_3$(100) substrate by molecular beam epitaxy. First, defect-related dI/dV tunneling spectra are identified by the K-means clustering method, followed by singular value decomposition to distinguish between those from S and Se. Such analysis provides an efficient and reliable determination of local elemental composition, and further reveals correlations of nanoscale chemical inhomogeneity to superconductivity in single-layer iron chalcogenide films.

## Introduction

Chemical doping is often exploited to tune material properties[1–5]. In semiconductors, aliovalent substitution leads to p- and n-type doping which is key for modern electronics[1]. Such substitution can also introduce chemical and electronic inhomogeneities at the nanoscale, which are often detrimental to carrier mobility and performance for semiconducting devices. On the other hand, for quantum materials, such heterogeneities are usually correlated and give rise to quantum phenomena such as superconductivity[3–5]. For example, most cuprate superconductors exhibit a significant amount of chemical disorder since isovalent or aliovalent substitutions that are substantially required to achieve superconductivity frequently lead to lattice site vacancies and/or interstitials. Traditionally, structural knowledge is often obtained using scattering techniques, where recent advances in x-ray scattering have enabled insight into chemical bonds with picometer resolution[6]. However, most such studies are inherently limited to macroscopically averaged properties. Alternatively, atomic scale imaging and spectroscopy by scanning tunneling microscopy (STM) and transmission electron microscopy (TEM) have been key to deciphering chemical disorder and electronic heterogeneity in quantum materials[7–9]. Nevertheless, traditional methods of visual inspection of these images can be challenging for multi-component alloys, particularly beyond the dilute limit due to chemical disorders and electronic inhomogeneities. One way to address this challenge is to use computational methods/statistical analysis (e.g., Fast Fourier Transform) to analyze the data obtained from imaging experiments to identify patterns and correlations that may not be immediately apparent through visual inspection alone[10]. A recent approach is to apply machine learning (ML) algorithms to analyze STM/TEM images, where they can be trained to recognize patterns and correlations in the data[11]. Such an approach has been applied to identify hidden electronic orders in STM imaging of electronic quantum matter and to determine surface chemical bonding and adatom interactions[12–14].

Here, we determine S and Se concentration based on machine learning of STM imaging of single-layer (SL) FeSe$_{1-x}$S$_x$, which further reveals a strong correlation between nanoscale chemical inhomogeneity and superconductivity. Iron chalcogenides exhibit a layer structure consisting of a plane of iron atoms with chalcogen (S, Se, Te) residing alternately above and below the plane, as sketched in Figures 1(a-b). Such structure enables the growth of SL FeSe film on SrTiO$_3$(001) (STO), which has the highest superconducting transition temperature in iron-based materials to date[15]. Nonetheless, the mechanism for Cooper pairing in such a system is still under debate, with several theories proposed including plain s-wave driven by electron-phonon coupling and d-wave driven by antiferromagnetic spin fluctuations[16–22]. To address this issue, we have shown that by substituting S into SL FeSe/STO, chemical pressure can be applied to tune its paramagnetic ground state and superconductivity[23]. In this approach, the local chemical composition is a critical parameter that can be obtained by atomic resolution STM imaging by visual inspection. However, chemical disorders and electronic inhomogeneities make such determination challenging beyond the dilute limit.

To determine the Se/S ratio for films with intermediate alloy compositions, ML methods were employed to analyze current imaging tunneling spectroscopy (CITS) maps in two steps. First, we identify the defect-related features based on analysis of dI/dV tunneling spectra using the supervised K-means method. Next, the remaining defect-free tunneling spectra are sorted into Se and S groups using the unsupervised singular value decomposition (SVD) method, based on which the Se/S ratio is calculated. Such analysis also reveals correlations between nanoscale chemical inhomogeneity and superconductivity. Our findings demonstrate an effective and reliable approach to determining alloy composition in single-layer superconducting iron chalcogenide films.

## Results and discussion

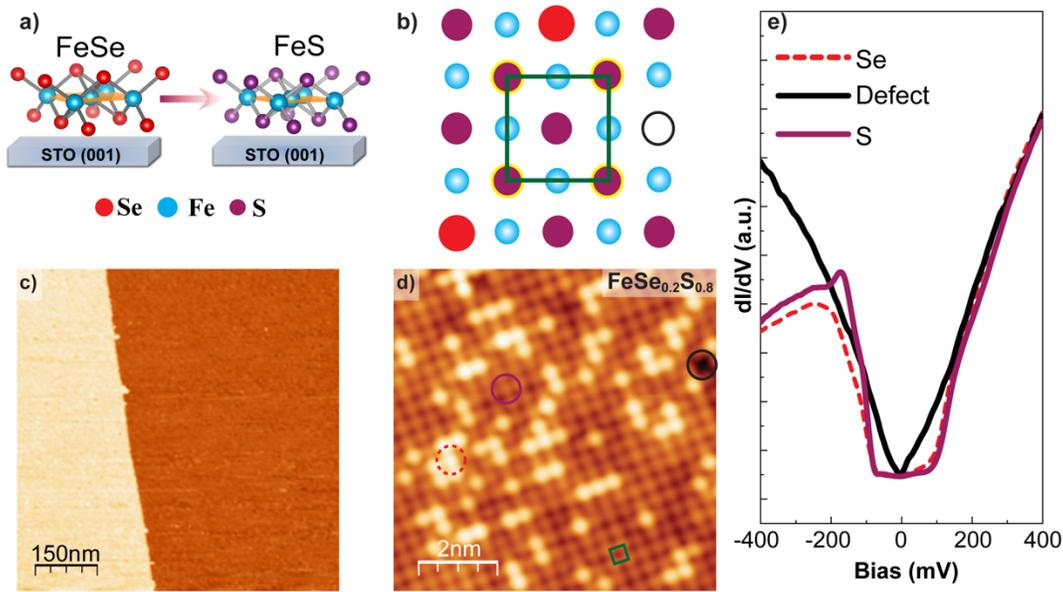

**Fig 1| STM image and tunneling spectra of single-layer FeSe$_{0.2}$S$_{0.8}$ on SrTiO$_3$(001) substrate.** (a) Ball-and-stick model of isovalent substitution of S into single-layer FeSe on STO substrate. (b) The top-view schematic of single-layer FeSe$_{1-x}$S$_x$ alloy (open circle represents a Se/S vacancy defect). (c) Large-scale and (d) atomic resolution STM image of single-layer FeSe$_{0.2}$S$_{0.8}$/STO. Imaging conditions: 3 V, 100 pA for (c) and 20 mV, 500 pA for (d). The green square in (d) marks the unit cell, and the black solid circle marks a Se/S vacancy). (e) Tunneling spectra taken at the three locations marked in (d). Set point: $V_{Bias}$ = 400 meV, $I_T$ = 500 pA.

The surface morphology of single-layer (SL) FeSe$_{1-x}$S$_x$/STO films was examined by STM measurements, where it is conformal to the step-terrace topography of the STO substrate (Figure 1c). At the diluted limit up to ~20% of Se (or S), their positions can be identified by the conventional visual inspection of atomic resolution STM images and dI/dV tunneling spectra. For example, because of their larger size, the Se atoms appear with higher contrast in STM images as marked by the red dashed circle in Figure 1d. The difference in Se and S can also be found in the dI/dV tunneling spectra taken as shown in Figure 1e. While the spectra acquired at Se and S atoms are both U-shaped near the Fermi level, the valence band edge shifts from -210 meV for Se to -180 meV for the S. By counting the numbers of Se and S atoms in the STM image, we determined that the Se concentration is 20 % for this film (see Fig. S1, SI for details and additional example). Note that a common type of defect observed is Se/S vacancy, as marked by the black circle in Figure 1d. Such a defect also caused a change of dI/dV tunneling spectra to be V-shaped for the whole energy range between -400 to 400 meV, which is dramatically different from those of Se and S sites.

The above method of visually inspecting atomically resolved STM images, which works for determining the Se/S concentrations in SL FeSe$_{1-x}$S$_x$/STO films in the diluted limit (see Fig. S1, SI for another example), becomes increasingly challenging for concentration, x, between 0.3 and 0.7 because the higher density of defects and electronic inhomogeneity present in the STM imaging of these films. This is illustrated in Figure 2a for the intermediate concentration case FeSe$_{0.43}$S$_{0.57}$, where the Se sites are hard to distinguish from those of S because of various sites with multiple high and low contrasts, which is also bias-dependent (see Fig. S2, SI for bias-dependent STM images). To determine the Se/S ratio in such a complex system, we apply ML to analyze CITS maps in two steps: 1) identify the defect-related tunneling spectra in the CITS map by the K-means clustering method, a supervised vector quantization data mining approach that groups a large dataset into components with principal characteristics[24]; 2) sort out the remaining (defect-free) CITS map into Se and S groups by the singular value decomposition (SVD) method[25,26].

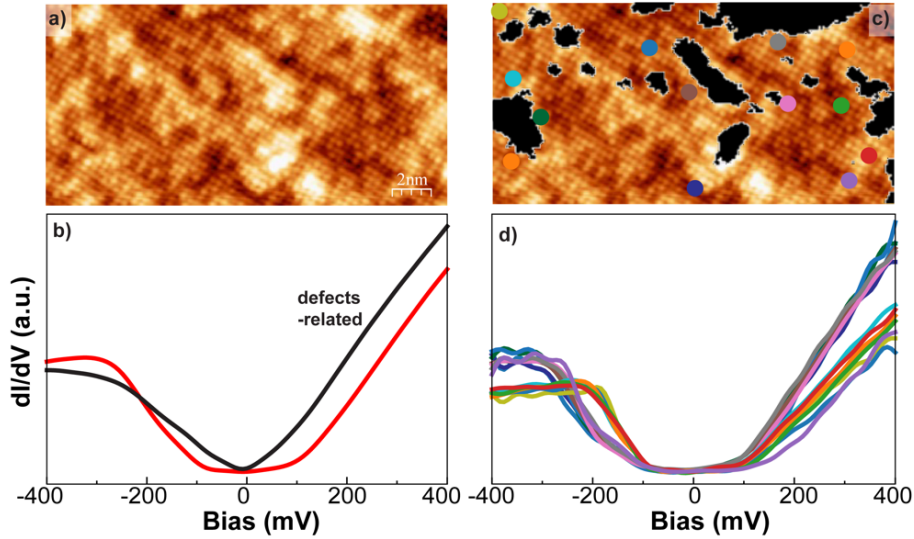

**Fig 2| K-means clustering analysis of defects in single-layer FeSe$_{0.43}$S$_{0.57}$.** (a) Atomic-resolution image of single-layer FeSe$_{0.43}$S$_{0.57}$. Imaging conditions: 100 mV, 100 pA. (b) Average tunneling spectra of the two clusters from K-means clustering of current imaging tunneling spectroscopy in (a). (c) Clusters map from K-means plotted over the atomic resolution image. (d) Tunneling spectra at the spots marked in (c). Set point: $V_{Bias}$ = 400 meV, $I_T$ = 500 pA.

In the first step, for the film shown in Figure 2a, tunneling spectra in the CITS map are cataloged into two groups, as shown in **Figure 2b**. The first group displays the characteristic V-shaped tunneling spectrum of defects, and the second group the U-shaped (Details of the K-means clustering are provided in SI Fig. S3). Then, the spatial distribution of those two types of tunneling spectra is overlaid on the atomic-resolution STM image, which reveals that the V-shaped spectra are located primarily at the very dark and the most bright locations. This analysis indicates that the defect regions are associated with V-shaped dI/dV tunneling spectra, and thus are excluded from the calculations of the Se/S ratio. Next, dI/dV are selected randomly in the rest of the CITS map as shown in Figure 2d. Interestingly, those tunneling spectra merge into two groups, characteristic of these for Se and S for films in the diluted limit (c.f., FeSe$_{0.2}$S$_{0.8}$ in Figure. 1e).

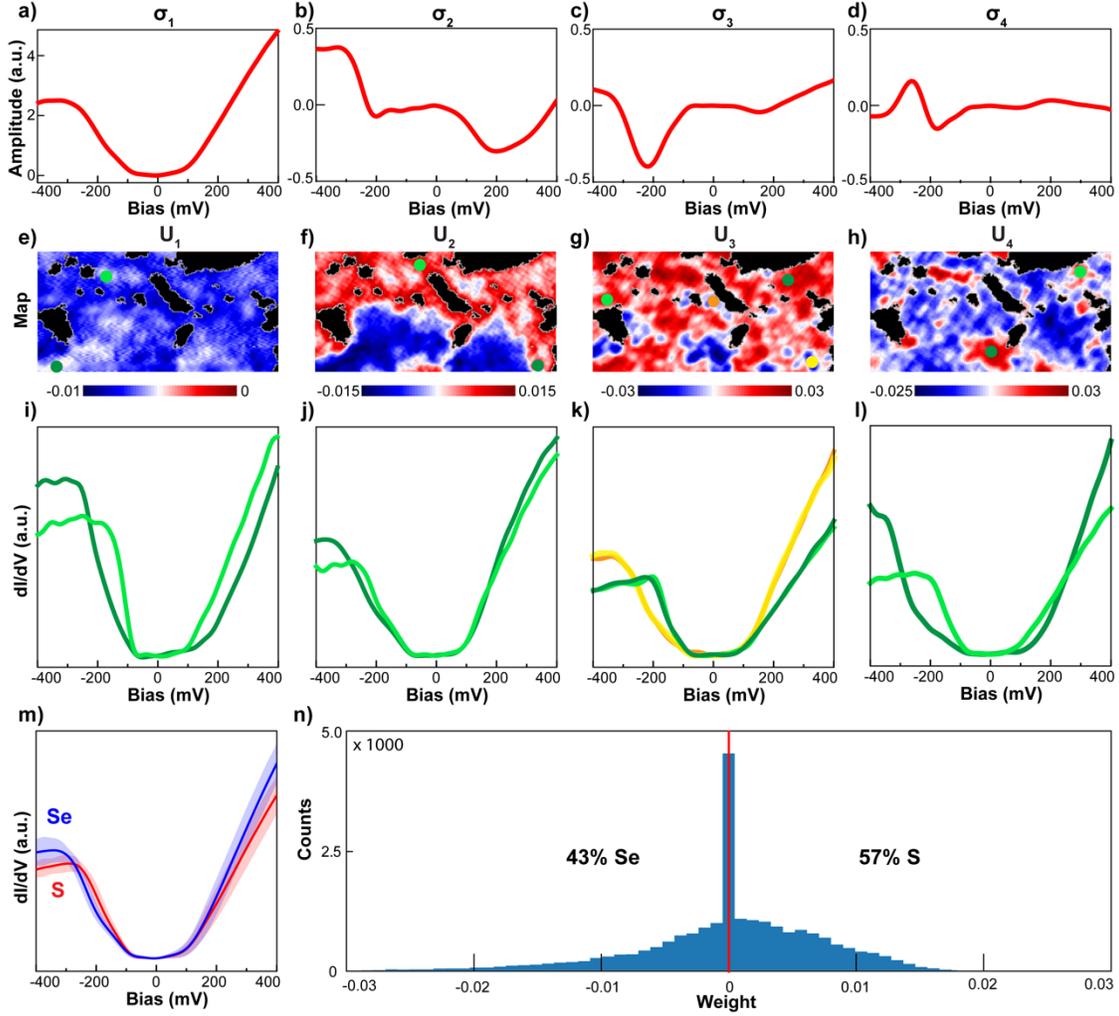

**Fig 3| Determination of the Se (S) concentration in single-layer FeSe$_{0.43}$S$_{0.57}$.** (a) and (e) ((b) and (f), (c) and (g) and (d) and (l)) are the $\sigma_{i=1, 2, 3, 4}$ and corresponding spatial distribution map U$_{i=1, 2, 3, 4}$, obtained by singular value decomposition, respectively. (i-l) Tunneling spectra were taken at the spots indicated in (e-h). (m) The average tunneling spectra with standard deviations at the red and blue locations of map U$_3$. (n) The statistics of spatial distribution map U$_3$.

To quantitively determine the Se and S concentrations, we utilize the SVD method to decompose the remaining tunneling spectra. The SVD is one of the most important matrix factorizations in computation, providing a numerically stable matrix decomposition and the optimal low-rank approximation for high-dimensional data[26]. Briefly, the optimal rank-r approximation to the matrix A is described by $\tilde{A} = \sum_{k=1}^{r} \sigma_1 U_1 V_1^T + \sigma_2 U_2 V_2^T + \cdots + \sigma_r U_r V_r^T$, where $k \in (1, 2, \ldots r)$ is the significant rank index, the columns of U ($\{U_i\}$) are called the left singular vectors, the columns of V($\{V_i\}$) are called the right singular vectors, and the {$\sigma_i^2$} ($\sigma_1 > \sigma_2 \ldots > \sigma_r$) are the eigenvalues of AA$^T$ (more details of the SVD method is provided in SI). For the film FeSe$_{0.43}$S$_{0.57}$, the four leading square root of eigenvalues ($\sigma_{i=1, 2, 3, 4}$) and their corresponding spatial distribution maps ($\{U_{i=1, 2, 3, 4}\}$) are shown in Figures (a-d) and (e-h), respectively (additional square root of eigenvalues and their spatial distribution maps are presented in SI. Fig. S4).

Close examination of the distribution maps reveals that the spatial distribution map $U_3$ successfully represents the spatial distribution of Se and S. Although the first leading square root of eigenvalue $\sigma_1$ (Figure 3a) is U-shaped, the corresponding spatial distribution map $U_1$ mixes the Se with S groups. Two spots are randomly selected from the same white color scheme in $U_1$, as indicated by the bright and dark green points in Figure 3e, then their corresponding tunneling spectra from the raw CITS dataset are plotted in Figure 3(i). The two tunneling spectra are clearly different, indicating that the $U_1$ doesn't distinguish Se from S. Similar failure is also found for other spatial distribution maps, such as the $U_2$ and $U_4$. As shown in Figures 3f-h, the two tunneling spectra in the red color scheme in each $U_{2, 4}$ are distinctive. In contrast, for $U_3$ (Figure 3g), the tunneling spectra at dark and bright green points in the red color scheme are mostly identical (Figure 3k). Moreover, the tunneling spectra at yellow and orange spots in the blue color scheme also overlapped. Those observations confirm that the $U_3$ successfully represents the spatial distribution of Se and S. It also indicates that the $\sigma_3$ captures the difference in tunneling spectra between Se and S in this sample.

We further confirm the successful grouping of $U_3$ by checking spatially dependent tunneling spectra presented in SI Fig. S5. The evolution of tunneling spectra from Se (U-shaped) to defect-related (V-shaped) to S (U-shaped) region is consistent with the $U_3$ map. Therefore, we concluded that the left singular vector $U_3$ displays the spatial distribution of Se in the blue color scheme and S in the red color scheme. The average tunneling spectra with standard deviations from Se and S atoms based on $U_3$ are presented in Figure 3m. Furthermore, based on the histogram of the spatial distribution $U_3$ (Figure. 3n), we determined the concentration of Se (S) is 57% (43 %). This ML method was also applied successfully to determine the alloy composition of other SL FeSe$_{1-x}$S$_x$/STO (results for FeSe$_{0.51}$S$_{0.49}$ are presented in SI Fig. S6).

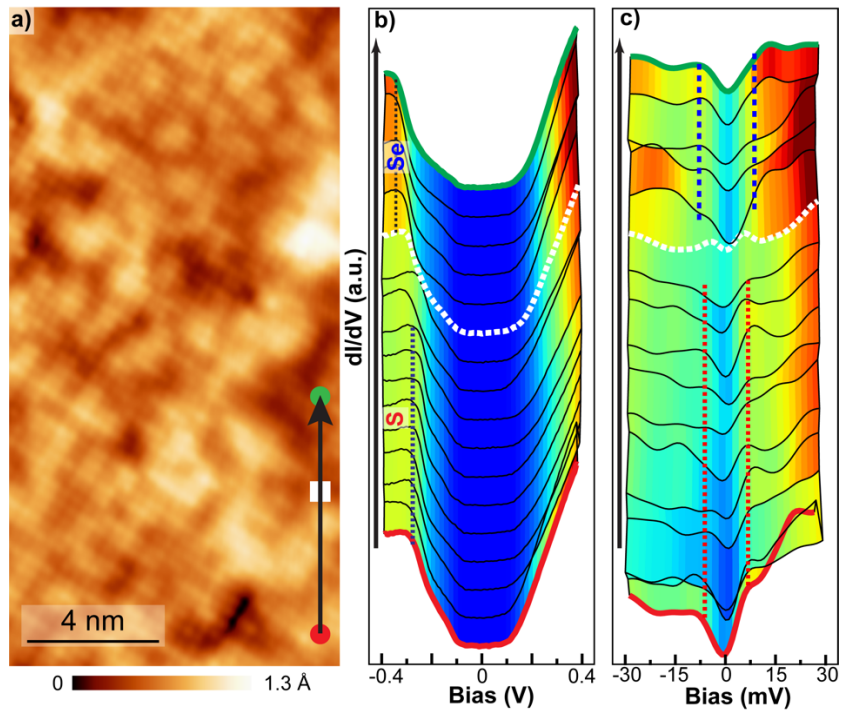

**Fig. 4 | Line STS in single-layer FeSe$_{0.43}$S$_{0.57}$.** (a) Atomic resolution image. (b-c) Line STS along the black arrow in (a) in a large bias range and around the Fermi level, respectively. Set point: $V_{Bias}$ = 400 meV, $I_T$ = 500 pA and $V_{Bias}$ = 30 meV, $I_T$ = 500 pA. T = 4.3 K.

The above analysis shows that the electronic structure of SL FeSe$_{1-x}$S$_x$/STO films exhibits high spatial inhomogeneity. Importantly, this inhomogeneity impacts the superconductivity locally, as indicated by the variation of the amplitudes and positions of superconducting coherence peaks in tunneling spectra shown in Figure 4c. Based on dI/dV tunneling spectra taken along the black line in Figure 4a, the valence band edge varies from -266 meV at S atoms to -327 meV at Se atoms, consistent with the results of Figure 3. Near the Fermi level, one pair of superconducting coherence peaks are clearly observed, while the gap values are correlated with the Se/S concentration. For the S-rich regions, the gap is smaller with an average of $\Delta_2 = 6.5 \pm 1$ meV, while the Se-rich regions exhibit a slightly larger value of $\Delta_1 = 8.0 \pm 1$ meV. Hence, the superconductivity of iron chalcogenide is tuned by the chemical pressure by the S substitution.

## Conclusion

High-quality single-layer FeSe$_{1-x}$S$_x$ films are epitaxially grown on SrTiO$_3$(001) using MBE and investigated using *in situ* low-temperature scanning tunneling microscopy/spectroscopy. For films at the diluted limit, the Se/S ratio is determined by the traditional visual inspection of atomic resolution STM images and dI/dV tunneling spectra. However, for films with intermediate alloy compositions where such an approach is challenging, we combine unsupervised with supervised machine learning methods on the spatially-dependent tunneling spectra to determine the chemical compositions. This ML-based approach provides an effective and reliable method to determine alloy concentrations and further reveal strong correlations between nanoscale chemical inhomogeneity and superconductivity in single-layer iron chalcogenide films.

## Methods

**Molecular beam epitaxy growth:** Single-layer FeSe$_x$S$_{1-x}$ films were grown on Nb-doped (0.5% wt) SrTiO$_3$ (001) substrates using molecular beam epitaxy. The SrTiO$_3$ (001) substrates were annealed at 1050 °C for 1 hour in an ultrahigh vacuum (UHV) MBE chamber (base pressure ~1 × 10$^{-9}$ Torr). After that treatment, the SrTiO$_3$ (001) surfaces show atomically flat terraces and sharp step edges, confirmed by *in-situ* STM. The growth of single FeSe$_x$S$_{1-x}$ films was carried out on the annealed SrTiO$_3$ (001) substrates. Fe was supplied using an electron beam source, and Se/S two separate Knudsen cells. The growth rate was about 0.5 monolayer/min, which was controlled by the Fe flux. The evaporating temperature of the Se source was kept at 106 °C during growth, and the S concentration was tuned by adjusting the evaporating temperature of the S source from 300 to 400 °C. The flux ratio of Fe:Se(S) is estimated to be 1:10-20. During the growth, the substrates were at 250 °C, and as-grown samples were further annealed at about 450 °C for 1 hour to achieve superconductivity.

**Scanning tunneling microscopy/spectroscopy:** The scanning tunneling microscopy and spectroscopy (STM/S) measurements were carried out in a Unisoku ultra-high vacuum low-temperature STM SPM-1300 system with a base temperature of 4.3 K. A polycrystalline PtIr tip was used and tested on Ag/Si(111) films before the STM/S measurements. Tunneling spectra were acquired using a standard lock-in technique with a small bias modulation V$_{mod}$ (2% of the setting point bias) at 732 Hz.

**K-means:** K-means method groups a large dataset into components with quintessential characteristics. The K-means clustering analysis uses a supervised learning algorithm to find similar groups in the data, where the number of groups is represented by the variable K. We used the same K-means codes as in reference [24].

**Singular value decomposition (SVD):** SVD is one of the most widely used multivariate statistical techniques for matrix decomposition. The purpose of singular value decomposition is to reduce a dataset containing a large number of values to a dataset containing significantly fewer values, which can still capture a large fraction of the variability present in the original data. We used the numpy.linalg.svd module to perform SVD analysis[27].

## Supplementary information

SVD method

Supplementary Figures S1–S6.

## Author contributions

L.L. conceived and organized the study. Q.Z., B.D.O., H.Z., and J. B. performed the MBE growth and STM/S measurements. Q.Z. and X.L wrote the codes for the machine learning. Q.Z., B.D.O., and L.L. analyzed the data and wrote the paper. All the authors read and commented on the paper.

## Notes

The authors declare no competing financial interest.

## Acknowledgments

This work is supported by the U.S. Department of Energy, Office of Basic Energy Sciences, Award No. DE-SC0017632 and DE-SC0021393.